\documentclass[journal,transmag]{IEEEtran}

\usepackage{cite}
\usepackage{amsmath}
\usepackage{booktabs}
\usepackage{tabularx}
\usepackage{subcaption}
\usepackage{import}
\usepackage{bm}
\usepackage{amsfonts}
\usepackage{amssymb}
\usepackage{soul}
\usepackage{graphicx}
\usepackage{xcolor}
\newcommand\myeq{\mathrel{\overset{\makebox[0pt]{\mbox{\normalfont\tiny\sffamily 1-D}}}{=}}}
\newcommand\myeqq{\mathrel{\overset{\makebox[0pt]{\mbox{\normalfont\tiny\sffamily 2-D}}}{=}}}

\begin{document}
	
	\title{Thin-Shell Approach for Modeling Superconducting Tapes in the $H$-$\phi$ Finite-Element Formulation}
	\author{\IEEEauthorblockN{Bruno de Sousa Alves\IEEEauthorrefmark{1}, Valtteri Lahtinen\IEEEauthorrefmark{2},
			Marc Laforest\IEEEauthorrefmark{1}, and
			Fr\'ed\'eric Sirois\IEEEauthorrefmark{1}}
		\IEEEauthorblockA{\IEEEauthorrefmark{1} Polytechnique Montr\'eal, Montr\'eal, QC, Canada}
		\IEEEauthorblockA{\IEEEauthorrefmark{2} QCD Labs, Department of Applied Physics, QTF Centre of Excellence, Aalto University, 00076, Aalto, Finland}
	}

	\IEEEtitleabstractindextext{
		\begin{abstract}
			This paper presents a novel finite-element approach for the electromagnetic modeling of superconducting coated conductors {with transport currents}. We combine a thin-shell (TS) method to the $H$-$\phi$-formulation to avoid the meshing difficulties related to the high aspect ratio of these conductors and reduce the computational burden in simulations. The interface boundary conditions in the TS method are defined using an auxiliary 1-D finite-element (FE) discretization of $N$ elements along the thinnest dimension of the conductor. This procedure permits the approximation of the superconductor's nonlinearities inside the TS in a time-transient analysis. Four application examples of increasing complexity are discussed: (i) single coated conductor, (ii) two closely packed conductors carrying anti-parallel currents, (iii) a stack of twenty superconducting tapes and a (iv) full representation of a HTS tape comprising a stack of thin films. In all these examples, the profiles of both the tangential and normal components of the magnetic field show good agreement with a reference solution obtained with standard $2$-D $H$-$\phi$-formulation. Results are also compared with the widely used $T$-$A$-formulation. This formulation is shown to be dual to the TS model with a single FE ($N=1$) in the auxiliary 1-D systems. The increase of $N$ in the TS model is shown to be advantageous at small inter-tape separation and low transport current since it allows the tangential components of the magnetic field to penetrate the thin region. The reduction in computational cost without compromising accuracy makes the proposed model promising for the simulation of large-scale superconducting applications.
			\vspace{-2mm}
		\end{abstract}
		
		\begin{IEEEkeywords}
			Finite-element method, high-temperature superconductors, thin-shell approach, transient analysis.
			\vspace{-4mm}
	\end{IEEEkeywords}}

	\maketitle
	\IEEEdisplaynontitleabstractindextext
	\IEEEpeerreviewmaketitle

	\section{Introduction}
	
	High-Temperature Superconducting (HTS) tapes are used in increasingly complicated geometries~\cite{Terzieva2010,Goldacker2014,wang2019,Hartwig2020} and the accurate simulation of their current distribution in such geometries is being held back by (i) the high aspect ratio of the meshes required to resolve the interior of the tapes and (ii) the nonlinearity of the $E$-$J$ relationship. Despite the diversity of available formulations of Maxwell's equations, the development of accurate and efficient numerical models is also hampered by the limitations of each formulation~\cite{biro1999edge}. The purpose of this research is to develop a finite-element (FE) model to accurately and efficiently predict current distribution and losses in HTS tapes.  The techniques proposed in this research are shown to be valid for time-domain models, treating nonlinear materials (both HTS and ferromagnetic), and complex magnetic field profiles in the tapes. 

\begin{figure}[t]
		\centering
		\begin{subfigure}{0.24\textwidth}
			\centering
			\includegraphics[width=1.15\textwidth]{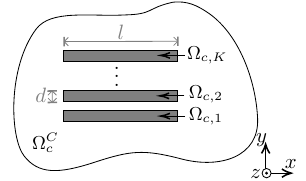}
			\caption{ Reference model}
			\label{domainFull}
		\end{subfigure}
		\begin{subfigure}{0.24\textwidth}
			\includegraphics[width=1\textwidth]{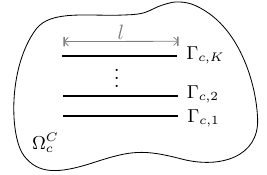}
			\caption{ TS model}
			\label{domainThin}
		\end{subfigure}
		\caption{ Computational domain comprising~$K$ disjoint thin conducting regions $\Omega_{c,1},\hdots \Omega_{c,K}$ and their representation in the TS model. The non-conducting region is denoted by $\Omega_c^C$ such that $\Omega=\Omega_c\cup\Omega_c^C$ and $\Omega_c=\cup_{i=1}^K \Omega_{c,i}$.}
		\label{domain}
		\vspace{-5mm}
	\end{figure}

	We assume a 2-D computational domain $\Omega$ comprising $K$ disjoint thin conducting regions 
	$\Omega_{c,1}, \ldots, \Omega_{c,K}$, each of which can be described in its own local coordinate system as
	a very long tape of length~$L$ in the $z$-direction (out-of-plane), of width \mbox{$l \approx 4$-$12$\,mm} in the $x$-direction, and of thickness $d \approx 1$\,$\mu$m in the $y$-direction (Fig.~\ref{domainFull}). The 
	current distribution is assumed constant in $z$. For HTS tapes, where the  magnetic field penetration generates sharp current fronts, traditional FE meshes require several elements in the $y$-direction and hence introduce elements with high aspect ratios. These irregular meshes lead to poor accuracy and slow convergence for
	iterative nonlinear solvers.
	
	In conventional (ohmic) conductors and linear ferromagnetic materials, the meshing of each thin layer can be avoided by using the classical \textit{thin-shell}~(TS) model~\cite{Mayergoyz1995,Krahenbuhl1993,Guerin1994,Biro1997}. In this case, the volume of the thin region (or its surface in a \mbox{2-D} case) is collapsed onto a surface (or a line in \mbox{2-D}) situated halfway in the \mbox{$y$-direction} between its upper and lower faces (Fig.~\ref{thin_domain}) while permitting discontinuities for field quantities across the surface. The TS model averages the magnetic flux density and current profiles in the transverse direction and couples the electric and magnetic effects by means of two interface conditions (ICs)~\cite{Biro1997}. The dimensionality reduction in this model avoids completely the need for meshing the thin region, resulting in gains in terms of computational cost~\cite{Bottauscio2004}. 
	
	The ICs in the classical TS model are obtained from the analytical solutions of the 1-D linear flux diffusion equation across the thickness of the thin structure~\cite{Biro1997}. Depending on the choice of the primary variables for the EM problem, these ICs can be written in terms of different field quantities~\cite{Geuzaine2000}. For example, in the weak form of the magnetic field \mbox{($H$-) formulation,} the interface terms arising from the application of Green's formula depend on the value of the \emph{tangential component of the electric field}~\cite{Bottauscio2006,Bottauscio2006b,Sabariego2008,Sabariego2009}, while in a magnetic vector potential \mbox{($A$-) formulation,} the equivalent interface terms depend on the \emph{tangential components of the magnetic field}~\cite{Sabariego2009,Sabariego2008,gyselinck2004time,gyselinck2008,Igarashi1998a,Igarashi1998b}. 
	
	Time-domain and nonlinear extensions of the TS model are usually derived from the ICs described above. In~\cite{gyselinck2004time}, the authors proposed using mathematical expansions in terms of Legendre polynomials to define ICs in laminated iron cores. The method is extended to nonlinear shielding analysis in~\cite{Sabariego2009}, and the nonlinear system of equations is solved using the Newton-Raphson~(NR) method. Both the $H$ and $A$ formulations are studied in that paper. In~\cite{Bottauscio2006}, the inverse fast Fourier transform of harmonic solutions with classical ICs is used to update the residual in nonlinear time-domain simulations. Shielding problems are also tackled in that paper, but only in the context of nonlinear ferromagnetic materials. So far, no time-domain TS model that considers tangential field penetrations has been proposed or applied to solve problems involving highly nonlinear conductive thin films, such as those found in HTS tapes.
	
	When the current density distribution across the thickness of the thin region is assumed constant, the ICs expressions can be simplified. This approach is called \textit{strip approximation}~\cite{Brandt1994}, and it has been originally implemented with integral methods~\cite{Grilli2009}. The strip approximation implies that the tangential components of $H$ have a linear profile inside the tape, which differs from the hyperbolic profiles encountered in the classical TS model~\cite{Alves2021}. Only the losses related to the normal flux density components (edge losses) are taken into account by this approximation, which means that the so-called top/bottom losses~\cite{Grilli2010} are disregarded. This is valid as long as the \emph{normal component} of $H$ dominates the dynamics of the problem, which is often the case when modeling simple problems involving HTS tapes, but not when these tapes are closely-packed together~\cite{Schacherer2017,Clem2008,Grilli2010b}. A strip approximation based on the current vector potential ($T$-formulation) and the Biot-Savart law was used to model HTS cables in~\cite{Takeuchi2011,Nii2012}, but this approximation does not scale well to large problems because of the nature of integral methods, which tend to produce full matrices.
	
	In recent years, a TS model based on a $T$-$A$-formulation and fully implemented in FE has been proposed~\cite{Zhang2016}. This model is now widely used to model HTS tapes~\cite{Liang2017,Berrospe2019,Benkel2020,Gao2020}. It involves solving an auxiliary \mbox{1-D} FE problem in terms of~$T$, related to the normal component of $H$ across the tape width. A surface current density is computed from $T$ and imposed in the standard \mbox{$A$-formulation}. However, combining the 1-D expressions for $T$ in the \mbox{$A$-formulation}, we find ICs similar to a TS model with linear $H$-profile inside the tape (or a constant current density across its thickness). As in the strip approximation, the \mbox{$T$-$A$-formulation} is only suitable for cases where the influence of the tangential components of $H$ can be neglected~\cite{Berrospe2019}. Moreover, the coupling between the state variables $T$ and $A$ requires higher-order FE for $A$ in order to avoid oscillations in the numerical solution~\cite{Berrospe2019}, which mostly limits its application to 2-D simulations~\cite{Yan2021}. A similar FE approach based on a mixed $A$-$H$-formulation has also been proposed \mbox{in~\cite{Bortot2020,Bortot2020b}}.
	
	In this paper, a new time-domain TS model is presented and applied to \mbox{a $H$}-based formulation for solving the current challenges in modeling HTS tapes. The ICs are here developed for 2-D cases and written in terms of a small number of auxiliary 1-D FE problems across the tape thickness to compute the penetration of the tangential components of~$H$. Four examples are presented: (i) single HTS tape, (ii) two closely packed tapes, (iii) a stack of twenty HTS tapes, and (iv) an HTS tape comprising three thin layers. The magnetic scalar potential ($\phi$) is considered in non-conducting parts of the computational domain, and edge-based cohomology basis functions (or thick cuts) are used to impose a net transport current in the tapes. Since the state variables in the 1-D and the global FE systems are naturally connected, shape functions of same order can be used in both the systems, and no numerical oscillations appear. Moreover, the use of a single element in the 1-D FE discretization makes the proposed model equivalent to the \mbox{$T$-$A$-formulation} under specific conditions, while the \mbox{1-D} mesh refinement enables improvements in the solution accuracy. We show that top/bottom and edge losses are correctly taken into account with our approach. All of these features make the proposed TS model ideal for simulating HTS devices in any context of use.

\vspace{-2mm}
	\section{Mathematical Formulation}
	
	Before deriving the TS model, we define the EM problem and review the so-called magnetic field and magnetic scalar potential \mbox{($H$-$\phi$)-formulation}, since our TS model is later integrated in this formulation. Note that the TS model presented in this paper could have also been rewritten in a pure $H$-formulation since $\phi$ is used mostly to reduce the computational cost in the non-conducting region.
	
	\vspace{-1mm}
	\subsection{Problem Definition}
	
	The electrodynamics of superconductors at low frequencies can be formulated as a nonlinear eddy-current problem. The computational domain is defined as \mbox{$\Omega = \Omega_c \cup	\Omega_c^C \subset \mathbb{R}^3$}, where $\Omega_c$ and $\Omega_c^C$ denote the conductive and non-conductive parts of $\Omega$, respectively. Moreover, $\Omega$ has a boundary $\partial \Omega$ denoted by $\Gamma$. The superconducting region, denoted by $\Omega_s$, belongs to $\Omega_c$, i.e., $\Omega_s\subset \Omega_c$. Neglecting displacement currents, the Maxwell equations governing this problem are
	\begin{align}
	\nabla \times \bm{h} & = \bm{j}, \label{Ampere} \\ 
	\nabla \times \bm{e} & = -\partial_t \bm{b}, \label{Faraday} \\
	\nabla \cdot \bm{b} & = 0, \label{divergence}
	\end{align}
	where bold characters represent vectors, $\bm{h}$ is the magnetic field, $\bm{b}$ is the magnetic flux density, $\bm{e}$ is the electric field and $\bm{j}$ the current density. The operator $\partial_t$ represents the time derivative. Additionally, two constitutive relations connecting these four fields quantities are required, i.e.
	\begin{align}
	\bm{b} &= \mu \bm{h}, \label{RC1} \\
	\bm{j} &= \sigma \bm{e}, \label{RC2}
	\end{align}
	where $\mu$ is the magnetic permeability and $\sigma$ is the electrical conductivity. 
	
	The boundary conditions impose the tangential components of either $\bm{h}$ or $\bm{e}$ in the following manner. We assume that there are known tangential vector fields $\bm{f}_h$ and $\bm{f}_e$ such that for all~$t$,
	\begin{align}
	\bm{n} \times \bm{h}(\bm{x},t) & = \bm{f}_h(\bm{x},t), \qquad \forall \bm{x} \in \Gamma_h, \label{BCh}\\
	\bm{n} \times \bm{e}(\bm{x},t) & = \bm{f}_e(\bm{x},t), \qquad \forall \bm{x} \in \Gamma_e, \label{BCe}
	\end{align}
	where $\bm{x}$ is any point in $\mathbb{R}^3$ and $\Gamma$ is subdivided into two complementary components $\Gamma_h$ and $\Gamma_e$ (i.e. $\Gamma =\Gamma_h \cup \Gamma_e$ and $\Gamma_h \cap \Gamma_e = \varnothing$) where boundary conditions~\eqref{BCh} and~\eqref{BCe}  may be applied, respectively.

	The resulting system \eqref{Ampere}-\eqref{divergence} is nonlinear when $\mu$ and $\sigma$ depend on the fields quantities. For type-II superconductors, the nonlinear $E$-$J$ relation is often expressed by a power-law characteristic~\cite{Rhyner1993}, i.e.
	\begin{equation}\label{powerlaw}
	\rho (\bm{j}) = \frac{e_c}{j_c} \left(\frac{| \bm{j} |}{j_c}\right)^{n-1} ,
	\end{equation}
	where $\rho$ is the electrical resistivity ($\rho=1/\sigma$) and  the electric field criterion $e_c$ also determines the critical current density~$j_c$~\cite{Dular2020}. The power index~$n$ determines the steepness of the $E$-$J$ curve.
	
	\vspace{-3mm}
	\subsection{\mbox{$H$-$\phi$-Formulation} } \label{Hphi}
	
	The well-known $H$-formulation is obtained from the weak form of Faraday's law~\eqref{Faraday}. Let  
	$ {\bm{H}}({\rm{curl}},\Omega) = \{ {\bm{u}} : \Omega \to {\mathbb{R}}^3 \, | \, \| {\bm{u}}\| < \infty, \| \nabla \times {\bm{u}}\| < \infty, {\bm{n}} \times {\bm{u}} = {\bm{f}}_h \text{ over }\Gamma_h \}$ where $\| {\bm{u}}\|^2 = \int_\Omega \| \bm{u} \|^2 d{\bm{x}} $ is
	the $L^2$ norm, while ${\bm{H}}_0({\rm{curl}},\Omega)$ is the same space except with homogeneous boundary conditions
	$\bm{n} \times \bm{u} = 0$ over $\Gamma_h$. Following the usual process, the weak formulation is : 
	
	\noindent Find $\bm{h} \in \bm{H}(\rm{curl},\Omega)$ such that 
	\begin{equation} \label{h_form_weak}
	\begin{split}
	{\Big( {\rho {\rm{ }}\nabla  \times {\bm{h}},\nabla  \times {\bm{g}}} \Big)_{{\Omega}}} &+ {\partial _t}{\Big( {\mu {\rm{ }}{\bm{h}},{\bm{g}}} \Big)_\Omega } \\ &\quad\quad + {\Big\langle {{\bm{n}} \times {\bm{e}},{\bm{g}}} \Big\rangle _{{\Gamma _{e}}}} = 0
	\end{split}
	\end{equation}
	$\forall$ ${\bm{g}}$ $\in$  $\bm{H}_0(\rm{curl},\Omega)$, where ${\bm{g}}$ are test functions, $\bm{n}$ is the outward unit normal vector on $\Gamma$, and 
	${\left( { \cdot , \cdot } \right)_\Omega }$ and ${\left\langle \cdot , \cdot \right\rangle _\Gamma }$ respectively denote the volume and surface integrals over $\Omega$ and $\Gamma$ of the scalar product of their two arguments. The last term in~\eqref{h_form_weak} is required to impose Neumann boundary conditions~\eqref{BCe} on the complementary surface portion $\Gamma_e$ of $\Gamma$, for physical or symmetry purposes~\cite{Meunier2010}. Also, $\bm{h}\in \bm{H}(\rm{curl},\Omega)$ already takes into account BC~\eqref{BCh}
	along $\Gamma_h$.
	
	In a pure $H$-formulation, $\bm{h}$ and $\bm{g}$ are normally described with the help of Whitney edge elements~\cite{bossavit1988whitney}. However, it is known that expressing the magnetic field in terms of the magnetic scalar potential \mbox{($\bm{h}=-\nabla \phi$)} in~$\Omega_c^C$ reduces the total number of DoFs and avoids errors such as leakage currents in $\Omega_c^C$, which appear in the pure \mbox{$H$-formulation~\cite{Lahtinen2015,Stenvall_2014}}. In the so-called \mbox{$H$-$\phi$-formulation,} nodal elements are used in~$\Omega_c^C$ and edge elements are used in~$\Omega_c$. Since $\nabla(S^0)\subset S^1$, where $S^0$ and $S^1$ are respectively the nodal and the edge FE space, $\bm{h}$ and $\phi$ are naturally connected at the interface between $\Omega_c$ and $\Omega_c^C$~\cite{Dular1997}. {A variant of the~\mbox{$H$-$\phi$-formulation} is the so-called \mbox{$T$-$\Omega$-formulation}, where $T$ is the current vector potential defined in the conducting parts of the domain and $\Omega$ is the magnetic scalar potential (equivalent to $\phi$) defined over the whole computational domain. This formulation was used to simulate HTS tapes in~\cite{Amemiya1998}.}
	
	When $\Omega_c^C$ is multiply connected, the representation of $\bm{h}$ through $\phi$ is not enough to express Amp\`ere's law in this region, since \mbox{$\nabla \times \nabla \phi=0$} $\forall \phi$. Uniqueness is resolved by imposing discontinuities in $\phi$ along thin cuts in $\Omega_c^C$, whose values determine the currents in connected components of~$\Omega_c$~\cite{kotiuga1987}. However, manually defining these thin cuts can be tiresome for complex geometries comprising multiple conducting subdomains. Alternatively, thick cuts can be uniquely determined by requiring that they be dual to the set of closed loops $C_i$ around each independent conducting subdomain $\Omega_{c,i}$, $i=1,\hdots,K$, with the loops being constructed during the meshing process. The computation of these so-called cohomology basis representatives is inexpensive and implemented in Gmsh~\cite{Pellikka2013}.
	
	The general discrete expression for the magnetic field is
	\begin{equation}\label{h_representation}
	\bm{h} = \sum_{e\in\Omega_c}^{} h_e \bm{w}_e + \sum_{n\in\Omega_c^C}^{} -\phi_n \nabla w_n + \sum_{C_i\in \Omega_c^C}^{} I_i \bm{\psi}_i,
	\end{equation}
	where $\bm{w}_e$ are the vector basis functions (edge elements) of each edge $e$ in $\Omega_c$, $w_n$ are the nodal basis functions of each node $n$ in $\Omega_c^C$, and $\bm{\psi}_i$ are the edge-based cohomology basis functions dual to the set of closed loops~$C_i$  whose coefficients $I_i$ correspond to the value of the integrals of $\bm{h}$ over these loops~\cite{Pellikka2013}. The discontinuity of $\phi$ is taken into account by the functions $\bm{\psi}_i$ so the \mbox{$H$-$\phi$-formulation} obeys Amp\`ere's law everywhere in $\Omega$~\cite{dular1999natural}. 
	
	The time derivative in~\eqref{h_form_weak} can be approximated by a finite difference discretization~(e.g. implicit Euler, Runge-Kutta, etc). Then, substituting \eqref{h_representation} into \eqref{h_form_weak}, and applying the Galerkin weighted residuals method, one obtains a nonlinear system of equations that can be solved, for example, with the \mbox{NR-method.} 
	
\begin{figure} [t]
		\centering
			\begin{subfigure}{.48\textwidth}
			\centering
			\includegraphics[width=1\textwidth]{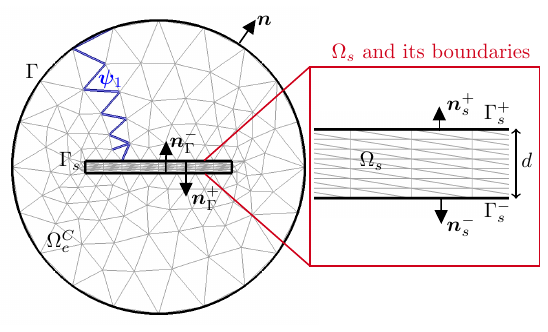}
			\caption{Fully discretized HTS tape for standard FE application}
			\label{full_domain}
		\end{subfigure}
		\begin{subfigure}{.48\textwidth}
				\centering
			\includegraphics[width=1\textwidth]{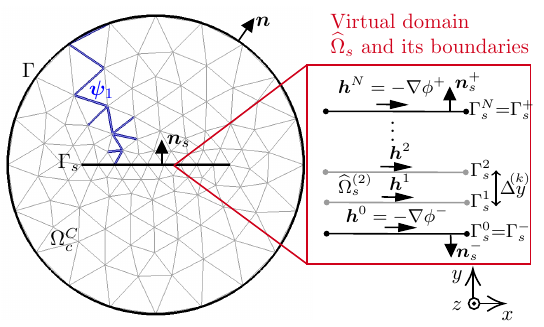}
			\caption{Proposed TS model for a HTS tape}
			\label{thin_domain}
		\end{subfigure}
		\caption{ (a) Fully discretized domain for standard FEM application and (b) proposed TS model: example of a single HTS tape embedded in a non-conducting domain $\Omega_c^C$ with boundary $\Gamma$. Nodes and edges on the surface representation of the thin structure in the TS model  are duplicated, creating a crack in the topological structure. The boundary of the thin region is $\Gamma_{s}=\Gamma_{s}^+\cup \Gamma_{s}^-$. The thick cut~$\bm{\psi}_1$ used to impose a current constraint in the tape is presented in blue. The 1-D FE system is directly connected to the global FE system of equations via Dirichlet conditions $\bm{h}_t^-=\bm{h}^0=-\nabla\phi^-$ in $\Gamma_{s}^-$ and $\bm{h}_t^+=\bm{h}^N=-\nabla\phi^+$ in $\Gamma_{s}^+$. The size of the air space region is not to scale, and the presented meshes are merely illustrative.}
		\label{Singletape}
		\vspace{-3mm}
	\end{figure}

	In this paper, the $H$-$\phi$-formulation is selected to construct the reference FE solutions and the TS model of the next section. In order to illustrate the geometry of~$\Omega$, Fig.~\ref{full_domain} shows a single HTS tape in an air space region, which is, in this case, a multiply connected region since the tape represents a hole in $\Omega_c^C$. A single thick cut~$\bm{\psi}_1$ is necessary to impose a current constraint in the tape and is presented in blue. The tape and the air space are not to scale. In the example, the mesh is defined with eight elements along the tape width and is merely illustrative. Note that for a fully discretized 2-D solution, elements with high aspect ratios are necessary inside the tape and near its extremities. A suitable discretization will be further considered to obtain reference solutions on a full-thickness geometry.
	
	\vspace{-2mm}
	\section{TS Model in the $H$-$\phi$-Formulation}
	
	In the TS model, $\Omega_s$ is collapsed to a surface~$\Gamma_{s}$ located  halfway between the original boundaries ($\Omega_{s}$ $\to$ $\Gamma_{s}$ in Fig.~\ref{thin_domain}).  Two surface integral terms are then included in the weak form of the $H$-formulation~\eqref{h_form_weak}, which becomes
	\begin{equation} \label{h_form_weak_int}
	\begin{split}
	&{\Big( {\rho {\rm{ }}\nabla  \times {\bm{h}},\nabla  \times {\bm{g}}} \Big)_{{\Omega}\setminus \Omega_s }} + {\partial _t}{\Big( {\mu {\rm{ }}{\bm{h}},{\bm{g}}} \Big)_{\Omega \setminus \Omega_s} } \\ 
	&\resizebox{.88\hsize}{!}{ $ + {\Big\langle {{\bm{n}} \times {\bm{e}},{\bm{g}}} \Big\rangle _{{\Gamma _{e}}}}
		-{\Big\langle {{\bm{n}_s} \times {\bm{e}},{\bm{g}}} \Big\rangle _{{\Gamma _{s}^+}}}
		+ {\Big\langle {{\bm{n_s}} \times {\bm{e}},{\bm{g}}} \Big\rangle _{{\Gamma _{s}^-}}}
		= 0 $},
	\end{split}
	\end{equation}
	where $\Gamma_{s}^+$ and $\Gamma_{s}^-$ are two copies of the collapsed boundary, at the same position but allowing independent variables along both boundaries. Note that the last two integral terms depend essentially on the tangential components of the electric field \mbox{$\bm{e}_t^+ = \bm{n}_s\times \bm{e} |_{\Gamma_{s}^+}$} and \mbox{$\bm{e}_t^- = \bm{n}_s\times \bm{e} |_{\Gamma_{s}^-}$,} where \mbox{$\bm{n}_s=-\bm{n}_\Gamma^+=\bm{n}_\Gamma^-$}. These terms are required to include the ICs in the TS model.
	
	The expression~\eqref{h_form_weak_int} requires the duplication of the DoFs on~$\Gamma_{s}$. Therefore, the surface mesh entities (nodes and edges on $\Gamma_s$) are duplicated, while the nodes at the extremities are not. This allows the tangential components of the magnetic field to be discontinuous across $\Gamma_{s}$. The thin surface corresponds to a crack in the topological structure of the discretized domain, and the non-conducting region becomes multiply connected. Consequently, a current constraint can be imposed in the thin structure using the shape functions $\bm{\psi}_i$ exactly as in the standard $H$-$\phi$-formulation. 
	
	\vspace{-2mm}
	\subsection{Interface Boundary Conditions  Derivation}
	
	To include the physics of the thin region in the simulations, the local subdomain $\Omega_{s}$ with boundaries $\Gamma_s^\pm$ is analyzed separately~(Fig.~\ref{full_domain}). The local weak form of the \mbox{$H$-formulation} in~$\Omega_s$ is written as
	\begin{equation} \label{Local}
	\begin{split}
	&-{\Big\langle {{\bm{n}_s} \times {\bm{e}},{\bm{g}}} \Big\rangle _{{\Gamma _{s}^+}}}
	+ {\Big\langle {{\bm{n_s}} \times {\bm{e}},{\bm{g}}} \Big\rangle _{{\Gamma _{s}^-}}} \\ 
	& \quad\quad\quad\quad\quad= {\Big( {\rho {\rm{ }}\nabla  \times {\bm{h}},\nabla  \times {\bm{g}}} \Big)_{{\Omega _s}}}  + {\partial _t}{\Big( {\mu {\rm{ }}{\bm{h}},{\bm{g}}} \Big)_{\Omega_s} },
	\end{split}
	\end{equation}
	where the integrals on~$\Gamma_{s}^\pm$ on the left side are identical to the two last terms in~\eqref{h_form_weak_int}. The lateral surfaces of $\Omega_{s}$ are neglected. This is valid since the thickness of the tape is much smaller than its width and length, and the field penetration from its wide faces can be assumed stronger than from its lateral faces, i.e., along its thickness.
	
	The domain integrals on the right side of~\eqref{Local} can be then reduced to boundary integrals on the collapsed geometry $\Gamma_{s}^\pm$, depending only on the tangential components of~$\bm{h}$ as follows.
	
	First, the profile of $\bm{h}$ in the $y$-direction across the thickness of the thin region is locally defined as a \mbox{1-D FE} problem in~$\widehat{\Omega}_s$ (Fig.~\ref{thin_domain}), where $\widehat{\Omega}_{s}$ is the virtual domain representing the original $\Omega_{s}$. The local coordinate systems $xyz$ is defined at the center of the thin geometry. The normal of the thin region is parallel to the \mbox{$y$-axis}, and $\bm{h}_t$ is in the \mbox{$x$-direction}. Since the tangential components of the magnetic field on $\Gamma_{s}^\pm$ are the same state variables as in~$\Omega_c^C$, the two subdomains~$\Omega_c^C$ and $\widehat{\Omega}_s$ are still naturally connected. For example, in Fig.~\ref{thin_domain}, the tangential fields in $\Gamma_{s}^+$ and $\Gamma_{s}^-$ are \mbox{$\bm{h}^N= \bm{h}_t^+ = -\nabla \phi^+$} and \mbox{$\bm{h}^0= \bm{h}_t^- = -\nabla \phi^-$,} respectively. 
	
	Then, by virtually discretizing $\widehat{\Omega}_{s}$ into $N$ FE of \mbox{size~$\Delta y^{(k)}$,} with $1\leq k\leq N$, the domain integral terms in~\eqref{Local} become
	\vspace{-0.5mm}
	\begin{equation} \label{elementary}
	\begin{split}
	{\Big( {\rho {\rm{ }}\nabla  \times {\bm{h}},\nabla  \times {\bm{g}}} \Big)_{{\Omega_s}}}
	&+ {\partial _t}{\Big( {\mu {\rm{ }}{\bm{h}},{\bm{g}}} \Big)_{\Omega_s}} \\
	& = \sum_{k=1}^{N}   {\Big( {\rho {\rm{ }}\nabla  \times {\bm{h}},\nabla  \times {\bm{g}}} \Big)_{{\widehat{\Omega}_s^{(k)}}}}\\
	&\quad\quad\quad + \sum_{k=1}^{N} {\partial _t}{\Big( {\mu {\rm{ }}{\bm{h}},{\bm{g}}} \Big)_{\widehat{\Omega}_s^{(k)}}},
	\end{split}
	\end{equation}
	where $\rho^{(k)}$ and $\mu^{(k)}$ are the electrical resistivity and magnetic permeability in element $k$ and
	the virtual domain $\widehat{\Omega}_s$ has been subdivided into $N$ slabs $\widehat{\Omega}_s^{(k)} = \Gamma_s \times [y_{k-1},y_k]$
	at heights $y_0 < y_1 < \cdots < y_N$.
	
	Using linear Lagrange polynomials as basis functions for the 1-D problem, i.e.
	\vspace{-0.5mm}
	\begin{equation} \label{Lagrange}
	\theta_1 (y) = \frac{y_{k}-y}{\Delta y^{(k)}}, \;\;\;\;\;\;\;\;\;\; \theta_2(y) = \frac{y-y_{k-1}}{\Delta y^{(k)}},
	\end{equation}
	we can write~$\bm{h}$ in each element~$\widehat{\Omega}_s^{(k)}$ as
	\vspace{-0.5mm}
	\begin{align}
	{\bm{h}} (x,y,t) \big|_{\widehat{\Omega}_s^{(k)}} = \sum_{j=1}^{2}{\bm{h}^{m}(x,t)} \theta_j(y), \label{hk}
	\end{align}
	where $m = m(k,j) =k-2+j$ relates $\bm{h}^{m}$ to the boundary of~$\widehat{\Omega}_s^{(k)}$, 
	i.e. $\bm{h}^{m(k,1)} = \bm{h}^{k-1}$ and $\bm{h}^{m(k,2)} = \bm{h}^{k}$. Recall that
	the vector field $\bm{h}$ is at all times perpendicular to the $y$-direction.
	
	Next, choosing the test function $\bm{g}(x,y)=\bm{g}(x)\theta_i(y)$, with $i=1,2$, to be in the same space as $\bm{h}$ (Galerkin method), we can use the vector calculus identity \mbox{$\nabla\times(\bm{v}u) = \nabla u \times \bm{v} + u\nabla\times \bm{v}$,} where $u$ and $\bm{v}$ are any scalar and vector functions, and decompose the domain integrals on~$\widehat{\Omega}_s^{(k)}$ in~\eqref{elementary} into integrals over $\Gamma_s^m$ and $[y_{k-1}, y_k]$.
	In fact, the decomposition below in 2-D uses the observation that $\nabla\times \bm{h}^{m}$ and $\nabla\times \bm{g}^n$ are in the
	$y$-direction and the inner products of type $(\nabla\theta_j\times \bm{h}^m)\cdot(\theta_i\nabla\times\bm{g}^n) $ vanish. Using these facts and setting $n=n(k,i)=k-2+i$, one deduces
	\color{black}
	\begin{equation} \label{term1}
	\begin{split}
	&{\Big( {\rho^{(k)} {\rm{ }}\nabla  \times {\bm{h}},\nabla  \times {\bm{g}}} \Big)_{{\widehat{\Omega}_s}}} \\
	&\quad\quad= \sum_{j=1}^{2} 
	\begin{split}
	\Big( \rho^{(k)} (\nabla\theta_j\times\bm{h}^m&+\theta_j\nabla\times\bm{h}^m), \\ &\mkern-18mu\mkern-18mu(\nabla\theta_i\times\bm{g}^n+\theta_i\nabla\times\bm{g}^n) \Big)_{\widehat{\Omega}_s^{(k)}} 
	\end{split}
	\\
	&\quad\quad\myeqq \sum_{j=1}^{2}  \Big\langle  \rho^{(k)} {\bm{h}}^{m},{\bm{g}^n} \Big\rangle_{\Gamma_s} \cdot \mathcal{S}^{(k)}_{ij} \quad\quad\quad \forall \, i=1,2,
	\end{split}
	\end{equation}
	and
	\begin{equation} \label{term2}
	\begin{split}
	&{\partial _t}{\Big( {\mu {\rm{ }}{\bm{h}},{\bm{g}}} \Big)_{\widehat{\Omega}_s^{(k)}} } \\
	&\quad\quad= \sum_{j=1}^{2} \partial_t \Big(\mu^{(k)} {\bm{h}}^{m} \theta_j, {\bm{g}^{m}} \theta_i \Big)_{\widehat{\Omega}_s^{(k)}} \\
	&\quad\quad=  \sum_{j=1}^{2} \partial_t \Big\langle  \mu^{(k)} {\bm{h}}^{m}, {\bm{g}^{n}} \Big\rangle_{\Gamma_s} \cdot \mathcal{M}^{(k)}_{ij} \quad \quad \forall \, i=1,2,
	\end{split}
	\vspace{-2mm}
	\end{equation}
	where
	\begin{equation} \label{Sij}
	\mathcal{S}^{(k)}_{ij} = \int_{\Delta y^{(k)}} \partial_y \theta_i \partial_y\theta_j dy, \;\;\;\;\; \mathcal{M}^{(k)}_{ij} = \int_{\Delta y^{(k)}} \theta_i \theta_j dy.
	\end{equation}
	
	Finally, the ICs in the proposed TS model are obtained by inserting~\eqref{term1} and~\eqref{term2} into~\eqref{elementary}. 
	Substituting~\eqref{elementary} in~\eqref{Local}, the interface terms depending on the tangential components of the electric field are written in terms of $\bm{h}^m$ as
	\color{black}
	\begin{equation} \label{SurfaceIntegralTerms}
	\begin{split}
	&-{\Big\langle {{\bm{n}_s} \times {\bm{e}},{\bm{g}}} \Big\rangle_{{\Gamma_{s}^+}}}
	+ {\Big\langle {{\bm{n_s}} \times {\bm{e}},{\bm{g}}} \Big\rangle_{{\Gamma_{s}^-}}} \\
	&\quad= \sum_{k=1}^{N}\sum_{j=1}^{2} \Big\langle  \rho^{(k)} {\bm{h}}^{m},{\bm{g}}^{n} \Big\rangle_{\Gamma_s^{m}} \cdot \mathcal{S}^{(k)}_{ij} \\
	&\quad+ \sum_{k=1}^{N}\sum_{j=1}^{2} \partial_t \Big\langle  \mu^{(k)} {\bm{h}}^{m}, {\bm{g}}^{n)} \Big\rangle_{\Gamma_s} \cdot \mathcal{M}^{(k)}_{ij}  \quad \forall \, i=1,2.
	\end{split}
	\end{equation}
	
	Given that polynomials of degree one were considered in~\eqref{Lagrange}, the profile of $\bm{h}$ in each element of~$\widehat{\Omega}_{s}$ is linear, and the current density~$\bm{j}_z$ is
	\begin{equation} \label{CurrentDensity}
	\bm{j}_z \Big|_{\widehat{\Omega}_s^{(k)}} = \bm{n}_s \times \partial_y {\bm{h}} \Big|_{\widehat{\Omega}_s^{(k)}} 
	= \bm{n}_s \times\left(\frac{\bm{h}^{k} - \bm{h}^{k-1}}{\Delta y^{(k)}}  \right),
	\end{equation}
	constant in~$\widehat{\Omega}_s^{(k)}$. Thus, the elementary 1-D \mbox{$E$-$J$ power-law~\eqref{powerlaw}} becomes
	\begin{equation}\label{powerlaw1D}
	\rho^{(k)} \myeq \frac{e_c}{j_c} \left(\frac{|\bm{n}_s\times ( {\bm{h}}^{k}-{\bm{h}}^{k-1} )|}{j_c \Delta y^{(k)}}\right)^{n-1}.
	\end{equation}
	Moreover, the matrices~\eqref{Sij} can be evaluated analytically for the polynomials~\eqref{Lagrange}, giving us
	\begin{equation} \label{Smat}
	\mathcal{S}^{(k)}= \frac{1}{\Delta y^{(k)}}
	\begin{bmatrix}
	1 & -1 \\
	-1 & 1
	\end{bmatrix}, \;\;\;
	\mathcal{M}^{(k)}= \frac{\Delta y^{(k)}}{6}
	\begin{bmatrix}
	2 & 1 \\
	1 & 2
	\end{bmatrix}.
	\end{equation}
	
	The complete weak form for the problem is obtained from~\eqref{h_form_weak_int} and~\eqref{SurfaceIntegralTerms}. 
	The time-derivative can be discretized by implicit Euler method, and the nonlinear system of equations is solved by the NR method.
	
	In terms of post-operation, the instantaneous loss density ($\mathcal{L}(t)$) in Joule inside the thin region is calculated as~\cite{Sabariego2009}
	\begin{equation} \label{AClosses}
	\mathcal{L}(t) =  \sum_{k=1}^{N} \int_{\Gamma_{s}} \ \rho^{(k)} H^{(k)T} \mathcal{S}^{(k)}  H^{(k)} d\Gamma,
	\end{equation}
	where $H^{(k)}$ is the $2\times1$ vector of unknowns for each element~$k$, i.e.
	\begin{equation}
	H^{(k)} = 
	\begin{bmatrix}
	h^{k} \\
	h^{k-1}
	\end{bmatrix},
	\end{equation}
	and $h^{k}$ and $h^{k-1}$ are the magnitude of the tangential field on $\Gamma_{s}^{k}$ and $\Gamma_{s}^{k-1}$, respectively, defined in~\eqref{hk}.
	
	The ICs in the proposed TS model were here developed using the polynomials of first order~\eqref{Lagrange}. For higher-order basis functions, \eqref{hk} must be modified accordingly. The approach can also be developed for the $A$-formulation. In this case, the boundary terms will depend uniquely on the tangential components of~$A$.
	
	\subsection{Comparison with the Classical TS model}
	
	For the sake of validation of the developed equations, we shall compare the ICs in the proposed model with those from the classical TS model for linear cases in harmonic \mbox{regime~\cite{Mayergoyz1995,Krahenbuhl1993,Guerin1994,Biro1997}.} Considering a single 1-D element in $\widehat{\Omega}_{s}$ ($N=1$), and $\delta\gg d$, where $\delta=\sqrt{2\rho/(\omega\mu)}$ is the skin depth and $\omega$ is the angular frequency, from~\eqref{SurfaceIntegralTerms}, one obtains
	\begin{align}
	\rho \frac{\bm{h}_t^+-\bm{h}_t^-}{d} &= \frac{\bm{e}_t^+ + \bm{e}_t^-}{2}, \label{IBC1}\\
	\frac{\bm{e}_t^+-\bm{e}_t^-}{d} &= -\partial_t \mu \frac{\bm{h}_t^++\bm{h}_t^-}{2} \label{IBC2},
	\end{align}
	which are identical to the ICs for $\delta\gg d$ presented in~\cite{Geuzaine2000}. 
	
	Expression~\eqref{IBC1} links the variation of $\bm{h}_t^\pm$ through the thickness~$d$ of the thin conductor (or the surface current density~\eqref{CurrentDensity}) to the mean value of $\bm{e}_t$, which is itself related to the mean value of the normal flux density by Faraday's law as
	\begin{equation} \label{faraday1D}
	\partial_x \bm{e}_t^\pm = -\mu\partial_t \bm{h}_n^\pm, 
	\end{equation}
	where $\bm{h}_n$ is the normal component of the magnetic field. Furthermore, expression~\eqref{IBC2} links the variation of $\bm{e}_t$ through $d$ (or the normal flux density) to the mean value of $\bm{h}_t$ (or the flux divergence on the thin surface). It expresses the conservation of the magnetic flux inside the thin region~\cite{Krahenbuhl1993}. 
	
	Depending on the problem one wishes to solve, the effects of the two ICs~\eqref{IBC1} and~\eqref{IBC2} are more or less predominant. For HTS modeling, with \mbox{$\mu=\mu_0$} and $\rho$ being highly nonlinear, \eqref{IBC1} becomes more important than~\eqref{IBC2}. Similar to the classical TS approach, the proposed model includes both ICs. However, since these ICs are defined from auxiliary 1-D FE problems, the TS model constructed in this paper permits time-transient and nonlinear analysis while considering the penetration of $\bm{h}_t$ in the thin region. 
	
	\vspace{-3mm}
	\subsection{Comparison with the $T$-$A$-formulation} \label{TAmodel}
	
	In the $T$-$A$-formulation, as presented and implemented in Comsol Multiphysics in~\cite{Zhang2016}, a 1-D FE problem problem is defined along the width of the thin regions. The state variable is the normal component of the current vector potential~($\bm{t}_n$), and the differential form for the 1-D problem is
	\begin{equation} \label{TA}
	\partial_x \big( \rho \partial_x \bm{t}_n \big) = - \partial_t\bm{b}_n
	\end{equation} 
	where $\bm{b}_n$ is the normal magnetic flux density. Moreover, the conventional $A$-formulation is applied to the air space region, and a surface current density~$\bm{j}_z$ is imposed on $\Gamma_s$ by a discontinuity in $\bm{h}_t$. The total transport current is imposed in the tapes via {Dirichlet} BCs of the type  $\bm{t}_n=I/{(2}d{)}$ at the extremities of~$\Gamma_{s}$.
	
	Since \mbox{$\bm{j}_z=\bm{n}_s\times\partial_x\bm{t}_n = \bm{n}_s\times(\bm{h_t}^+-\bm{h_t}^-)/d$,} and \mbox{$\bm{b}_n = (\bm{b}_n^+ + \bm{b}_n^-)/2$,} \eqref{TA} with~\eqref{faraday1D} corresponds to IC~\eqref{IBC1}. Also, in Comsol Multiphysics, the interface condition \mbox{$\bm{e^+}_t-\bm{e}_t^-=0$} is automatically fulfilled~\cite{Comsol}. So, IC~\eqref{IBC2} is not taken into account in the $T$-$A$-formulation. Since $\mu=\mu_0$ in~$\Omega_s$, this approximation is valid for modeling HTS tapes only when the superconducting layer is represented as a thin strip. Problems involving thin regions with $\mu>\mu_0$ cannot be addressed by this formulation since a conservative normal flux density is implicitly imposed by \mbox{$\bm{e^+}_t-\bm{e}_t^-=0$}.
	
Given that $T$-$A$-formulation can be defined simply as the IC~\eqref{IBC1} applied in a pure $A$-formulation, it is proven to be dual to the proposed TS model with $N=1$ in the $H$-formulation and when disregarding the magnetic flux divergence on the thin surface given by~\eqref{IBC2}. However, its use should be carefully considered. The current density depends at the same time on both the variation of~$\bm{h}_t$ across the thickness $d$ and the variation of~$\bm{h}_n$ along its width. The former variation is assumed linear, and the second must also be. From~\eqref{faraday1D}, one verifies that, for $\bm{h}_t$ to be linear, $\bm{e}_t$ must be of second-order. Otherwise, numerical oscillations appear~\cite{Berrospe2019,Yan2021}. Consequently, with second order FE for $\bm{e}_t$ (or $\bm{a}_t$), the number of DoFs greatly increases in the \mbox{$T$-$A$-formulation}. The computation time with this formulation becomes comparable to a pure $H$-formulation in 3-D cases~\cite{Yan2021}. 
	
	In the proposed TS model, the ICs are written only in terms of~$\bm{h}_t$, which are also the state variables in the \mbox{$H$-formulation} used to model the surroundings of the thin region. This represents a more natural coupling and avoids the oscillations that appears in the \mbox{$T$-$A$-formulation.} The application of the equivalent TS model in a pure \mbox{$A$-formulation} would require ICs depending on the tangential components of the electric field. However, this formulation might be preferable to model soft ferromagnetic materials. For superconducting tapes, the use of the $H$-formulation is more advantageous~\cite{Dular2020}. 
	
	\section{Validation}
	
	The proposed TS model was applied in two benchmark problems: (i) a single HTS tape, (ii) two closely-packed tapes carrying anti-parallel currents. In both examples, width of the tapes was $l=4$~mm, and their thickness was scaled to~$d=10$\,$\mu$m in the full-discretized reference model in order to reduce the number of DoFs while ensuring reasonable accuracy. Despite this, the aspect ratio of the tapes remains relatively high. Only the superconducting layer is modeled in these examples. The magnetic permeability is equal to $\mu_0$ in all domains.
	
	In the $E$-$J$ power-law model~\eqref{powerlaw}, and in~\eqref{powerlaw1D}, \mbox{$e_c = 10^{-4}$\,V/m,} \mbox{$j_c=5\times 10^8$\,A/m$^2$} and \mbox{$n=21$}. Moreover, a sinusoidal transport current of \mbox{$I(t) = I_{\text{max}} \sin (2\pi f t)$} is imposed in the tapes, where \mbox{$f=50$~Hz} is the operating frequency and \mbox{$I_{\text{max}} = F_c I_c$,} where \mbox{$I_c=J_c S$} is the critical current density, $S$ is the cross-section area of the tape, and $F_c\in [0,1]$ is a constant defining the transport current amplitude as a fraction of the critical current. 
	
	The results obtained with the proposed TS model are compared with reference solutions obtained with standard FE using the $H$-$\phi$-formulation presented in Section~\ref{Hphi}, and with the $T$-$A$-formulation proposed in~\cite{Zhang2016} and briefly described in Section~\ref{TAmodel}. In output, the local distribution of magnetic field inside and outside the tapes, current density, and total AC losses in the tapes are compared.
	
\begin{figure}[t]
			\begin{subfigure}{0.24\textwidth}
				\includegraphics[]{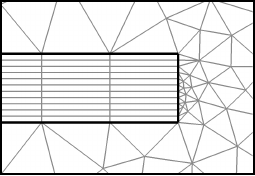}
				\caption{Full mesh}
				\label{meshfull}
			\end{subfigure}\hspace{0.5mm}
			\begin{subfigure}{0.24\textwidth}
				\includegraphics[]{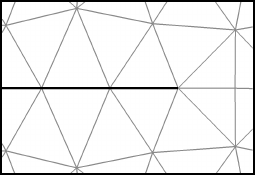}
				\caption{Mesh TS model}
				\label{meshthin}
			\end{subfigure} 
		\caption{ Zoom of the meshes near the right edge of the tape: (a) full mesh for standard FE application (reference model), and (b) simplified mesh used with the the TS model and the $T$-$A$-formulation. Note the difference in the number of elements surrounding the extremity of the tape and their aspect ratio.}
		\label{mesh}
		\vspace{-2mm}
	\end{figure}

	In terms of mesh, the HTS tapes  in the reference model are fully discretized with a structured rectangular mesh with 11 elements across the tape thickness and 400 elements along its width (Fig.~\ref{meshfull}). In the proposed TS model and in the \mbox{$T$-$A$-formulation,} the surface representing the thin region also includes 400 elements across its width (Fig.~\ref{meshthin}). The influence of the number of 1-D elements~($N$) in the TS model is studied independently in each application example. The total number of DoFs is further compared with the reference model.
	
	The reference and the TS models were implemented in the open-source code Gmsh~\cite{Gmsh} and the solver GetDP~\cite{Getdp}. The results of the $T$-$A$-formulation were obtained using Comsol Multiphysics 5.5. All the simulations were performed using a personal computer with an Intel i7 2400 processor with 16\,Gb of memory. In GetDP, an adaptive time-step procedure was used to improve convergence. This procedure is well explained in~\cite{Dular2020}. In our simulations, the NR scheme was used to solve the nonlinear system of equations. The convergence criterion requires  the relative change between two consecutive NR iterations to be smaller than a given tolerance. The maximum number of iterations was set to 12, the initial time step was set 7.5\,$\mu$s, the maximum time step was set 200\,$\mu$s. The relative tolerance of the calculation was set to $10^{-5}$, with absolute tolerance set to $10^{-7}$.  Moreover, a zeroth-order extrapolation was used, i.e. the initial guess in each \mbox{NR iteration} was taken as the solution of the previous time step.
	
	\subsection{Single HTS Tape} \label{Onetape}
	
	We first considered the example of a single HTS tape embedded in an air space region, as illustrated in Fig.~\ref{Singletape}. The origin of the coordinate system~$xy$ was located at the center of the tape. The number of \mbox{1-D elements}~($N$) was initially set to 1. The current imposed in the tape was 0.9$Ic$, i.e. $F_c=0.9$.
	
\begin{figure}
			\begin{subfigure} {0.24\textwidth}
			\centering
			\includegraphics[]{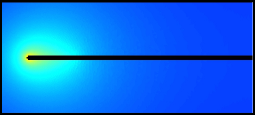}
			\caption{$H$-$\phi$ formulation ($t= T/8$)}
			\vspace{2mm}
		\end{subfigure}
		\begin{subfigure} {0.24\textwidth}
			\centering
			\includegraphics[]{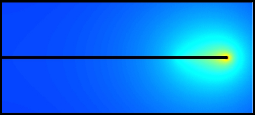}
			\caption{$H$-$\phi$ TS model ($t=T/8$)}
			\vspace{2mm}
		\end{subfigure}
		\begin{subfigure} {0.24\textwidth}
			\centering
			\includegraphics[]{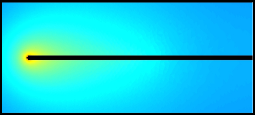}
			\caption{$H$-$\phi$ formulation ($t= T/4$)}
			\vspace{2mm}
		\end{subfigure} 
		\begin{subfigure} {0.24\textwidth}
			\centering
			\includegraphics[]{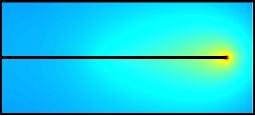}
			\caption{$H$-$\phi$ TS model ($t=T/4$)}
			\vspace{2mm}
		\end{subfigure} 
		\begin{subfigure} {0.24\textwidth}
			\centering
			\includegraphics[]{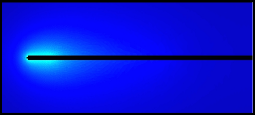}
			\caption{$H$-$\phi$ formulation ($t= T/2$)}
			\vspace{2mm}
		\end{subfigure}
		\begin{subfigure} {0.24\textwidth}
			\centering
			\includegraphics[]{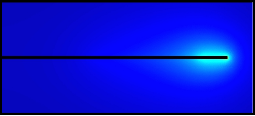}
			\caption{$H$-$\phi$ TS model ($t=T/2$)}
			\vspace{2mm}
		\end{subfigure}
		\begin{subfigure}{0.49\textwidth}
			\vspace{0.1cm}
			\includegraphics[]{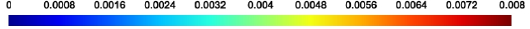}
			\label{scaleFig4}
		\end{subfigure}
	\vspace{-2mm}
		\caption{ Norm of magnetic flux density $|\bm{b}|$~[T] for the single HTS tape example with $F_c=0.9$ at three different simulation times ($T/8$, $T/4$ and $T/2$) . Left: fully discretized \mbox{$H$-$\phi$} reference model. Right: proposed TS approach with $N=1$.}
		\label{induction}
		\vspace{-1mm}
	\end{figure}

	In Fig.~\ref{induction}, the distribution of the magnetic flux density in and around the tape is shown for the TS model and for the $H$-$\phi$ reference model at $t=[T/8,T/4, T/2]$, where \mbox{$T=1/f$}. Due to symmetry, only half of the tape is shown. The figures on left show the flux distribution obtained with the $H$-$\phi$ model, and the figures on right show the results obtained with the TS model with $N=1$. The distributions computed by the two models show excellent agreement, validating the proposed TS model in terms of field distribution outside the tape.
	
\begin{figure} [t]
		\includegraphics[]{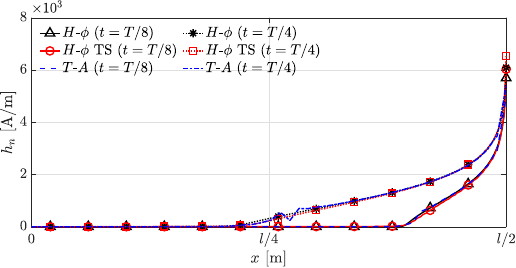}
		\caption{ Profile of normal component of $H$ ($h_n=|\bm{h}_n|$) along half of the tape width with $I=0.9I_c$ and at $t=T/8$ and $t=T/4$. The TS solution with $N=1$ is compared with the $H$-$\phi$ and $T$-$A$ solutions. }
		\label{MagneticField_Width}
	\end{figure}

	For the sake of comparison with the $T$-$A$-formulation and solution validation inside the tape, Fig.~\ref{MagneticField_Width} shows the profile of the normal component of $H$~(${h}_n=|\bm{h}_n|$) along half of the tape width \mbox{($0<x<l/2$),} and at $t=T/8$ and $t=T/4$. In the TS model ($H$-$\phi$ TS), the $h_n$ profile was obtained by computing its mean value on surfaces $\Gamma_{s}^\pm$, i.e., \mbox{${h}_n=({h}_n^++{h}_n^-)/2$.} In the reference model ($H$-$\phi$), $h_n$ was taken at the middle of the tape \mbox{($y=0$).} Moreover, in the \mbox{$T$-$A$-formulation,} \mbox{$h_n= (b_n^+ + b_n^-)/2\mu_0$,} with $b_n^\pm=|\bm{b}_n^\pm|$ derived from $A$. Note the excellent agreement of the $h_n$ profiles of the TS model with the two other solutions. The slight difference observed at the extremities of the tape may be related to the geometrical differences between the TS and the reference model. The thinner the thin film is, the lower the difference between the TS and the reference solutions will be at these extremities.
	
\begin{figure}[t]
		\includegraphics[]{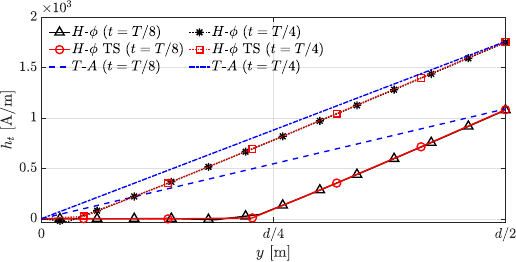}
		\caption{ Profile of tangential component of $H$ (${h}_t=|\bm{h}_t|$) across half of the tape thickness with $I=0.9I_c$ and at $t=T/8$ and $t=T/4$. The TS solution with \mbox{$N=11$} is compared with the $H$-$\phi$ and $T$-$A$ solutions. }
		\label{MagneticField_Thickness}
	\end{figure}

	The profile of the tangential component of $H$~(${h}_t=|\bm{h}_t|$) across the tape thickness at $x=0$ is presented in Fig.~\ref{MagneticField_Thickness}. In this case, $N=11$ to correspond to the number of mesh elements across the thickness in the fully discretized $H$-$\phi$ model. The profile of $h_t$ in the \mbox{$T$-$A$-formulation} was obtained by linear interpolation of the field intensities at the center of the tape. This solution is similar to the TS model solution with $N=1$. Indeed, with the increase in $N$, the proposed TS model allows the tangential components of the fields to penetrate the tape.
	
	In the modeling of a single HTS tape, the increase in $N$ in the TS model does not have a major impact on the quality of the solution. From Fig.~\ref{MagneticField_Width} and Fig.~\ref{MagneticField_Thickness}, one observes that $h_n$ is larger than $h_t$ ~($\sim3.5$ times). Thus, the assumption of a linear variation of $h_t$ across~$d$ is reasonable for a single tape, and $N=1$ gives sufficiently accurate solutions, similar as the \mbox{$T$-$A$-formulation}.
	
\begin{figure}[t]
			\begin{subfigure}{0.24\textwidth}
			\includegraphics[]{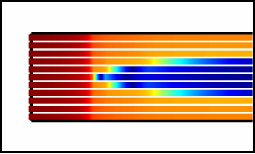}
			\caption{$H$-$\phi$ formulation ($t=T/8$)}
			\label{currentFullT8}
		\end{subfigure}\hspace{0.5mm}
		\begin{subfigure}{0.24\textwidth}
			\includegraphics[]{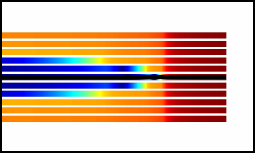}
			\caption{$H$-$\phi$ TS model ($t=T/8$)}
			\label{currentThinT8}
		\end{subfigure}\hspace{0.5mm}
		\begin{subfigure}{0.24\textwidth}
			\includegraphics[]{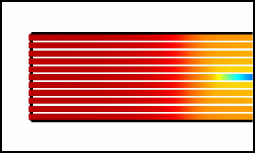}
			\caption{$H$-$\phi$ formulation ($t=T/4$)}
			\label{currentThinT4}
		\end{subfigure} 
		\begin{subfigure}{0.24\textwidth}
			\includegraphics[]{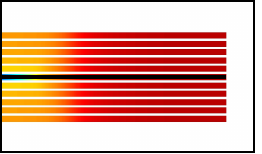}
			\caption{$H$-$\phi$ TS model ($t=T/4$)}
			\label{currentFullT4}
		\end{subfigure}
		\begin{subfigure}{0.24\textwidth}
			\includegraphics[]{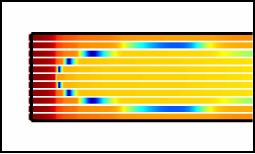}
			\caption{$H$-$\phi$ formulation ($t=T/2$)}
			\label{currentThinT2}
		\end{subfigure} 
		\begin{subfigure}{0.24\textwidth}
			\includegraphics[]{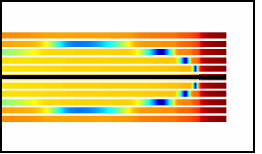}
			\caption{$H$-$\phi$ TS model ($t=T/2$)}
			\label{currentFullT2}
		\end{subfigure}
		\begin{subfigure}{0.49\textwidth}
			\vspace{0.1cm}
			\includegraphics[]{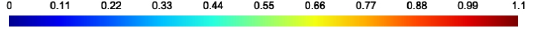}
			\label{scaleFig7}
		\end{subfigure}
		\vspace{-3mm}
		\caption{ Normalized relative current density ($|\bm{j}_z|/J_c$) near the extremity of the tape at $t=T/8$, $t=T/4$ and $t=T/2$: (a,c,e) $H$-$\phi$-formulation, and (b,d,f) $H$-$\phi$ TS model with $N=11$. The TS solution is obtained by a projection of the $|\bm{j}_z|/J_c$ on the virtual position of the 1-D elements. The black lines in the middle of (b,d,f) represent the tape in the TS model (lines in 2-D).
		}
		\label{Current}
	\end{figure}

	One advantage of the TS model with the discretization across the tape thickness is that the current density~$\bm{j}_z$ can be locally evaluated. Fig.~\ref{Current} shows the normalized current density relative to $j_c$ ($|\bm{j}_z|/j_c$) near the extremity of the tape for $N=11$. The results of the TS model (Fig.~\ref{Current}b,d,f) were obtained by a projection of $|\bm{j}_z|/j_c$ across~$d$, and coincides with the reference solution (Fig.~\ref{Current}a,c,e).
	
\begin{figure} [t]
		\centering	
		\includegraphics[]{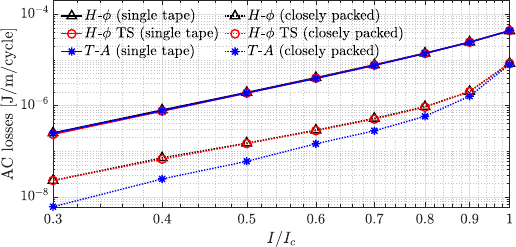}
		\label{Losses_percycle_closelyPacked}
		\vspace{-2mm}
		\caption{ Total AC losses as a function of the normalized transport current in the single tape and in the two closely packed tapes examples. For a single tape $N=1$, and, for two closely packed tapes, $N=11$ in the TS model.}
		\label{LossesComp}
		\vspace{-3mm}
	\end{figure}

	For the TS model, the AC losses inside the tape were evaluated using~\eqref{AClosses} with $N=1$. The results are presented in~Fig.~\ref{LossesComp} as a function of $I/I_c$. For a single tape, the losses from the three discussed models are similar. The differences in terms of AC loss appears when more than a one tape is studied.
	
	\subsection{Two Closely Packed HTS Conductors} \label{2CloselyPacked}
	
	In large-scale HTS applications, the use of more than one tape is required. When these tapes are closely packed and carry  anti-parallel currents, the tangential components of the magnetic field are important and the current density is unevenly distributed across the thickness of the tape. Consequently, the strip approximation and the \mbox{$T$-$A$-formulation} cannot provide an accurate solution, and a pure $H$-formulation requires tremendous computational efforts to discretize the thin region~\cite{Grilli2009}.  
	
\begin{figure} [t]
		\vspace{-3mm}
		\centering
		\includegraphics[width=0.4\textwidth]{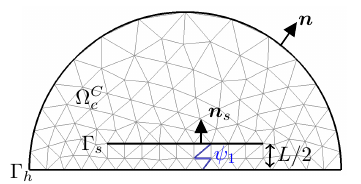}
		\caption{ Calculus domain for the example of two closely packed HTS tapes. Due to the symmetry, only half of the geometry is presented. The distance between the conductors is $L$. The air space region, the mesh and the distance $L$ are not to scale.}
		\label{GeometryCloselyPacked}
	\end{figure}

	To demonstrate that the proposed TS model is suitable for simulating HTS problems of any type, an example of two closely-packed tapes carrying anti-parallel currents was considered. The inter-tape separation was $L=250$\,$\mu$m. Only half of the geometry was modeled due to symmetry (Fig.~\ref{GeometryCloselyPacked}). One tape is located at a distance $L/2$ from the exterior boundary $\Gamma_h$. Since the BC on $\nabla\times\bm{h}$ along $\Gamma_h$ is naturally satisfied in the $H$-$\phi$-formulation, symmetry condition is automatically fulfilled. 
	
\begin{figure} [t]
		\centering	
		\includegraphics[]{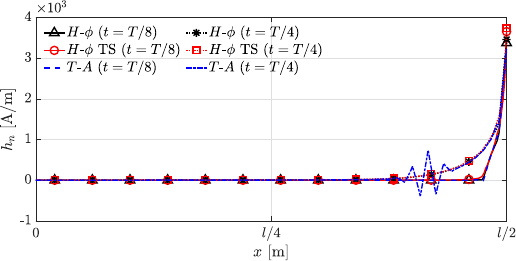}
		\caption{ Profile of normal component of $H$~($h_n=|\bm{h}_n|$) along half of the tape width for $I=0.9I_c$ and at $t=T/8$ and $t=T/4$ for the case of two closely packed tapes. The TS solution with $N=1$ is compared with the $H$-$\phi$ and $T$-$A$ solutions. }
		\label{MagneticField_Width_packed}
		\vspace{-4mm}
	\end{figure}

	Fig.~\ref{MagneticField_Width_packed} shows the profile of the normal component of $H$ (${h}_n=|\bm{h}_n|$) along half of the tape width. The solution of the TS model with $N=1$ is compared with the solutions of the $H$-$\phi$ and the $T$-$A$ models for $F_c=0.9$. Once again, the solution of the three models agree. In the $T$-$A$ solution, one observes oscillations at the sharpest points of the ${h}_n$ profile. No oscillations are noted in the TS solution.
	
\begin{figure} [t]
		\centering	
		\includegraphics[]{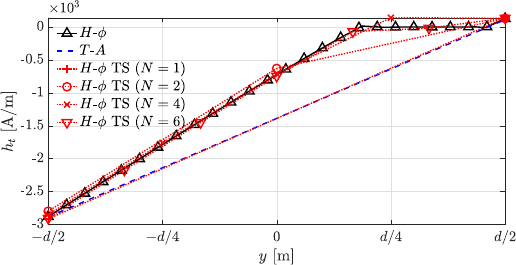}
		\caption{ Profile of tangential component of $H$~(${h}_t=|\bm{h}_t|$) across half of the tape thickness with $I=0.9I_c$, and at $t=T/8$ and $t=T/4$. The TS solution with \mbox{$N=1$, 2, 4 and 6} is compared with the $H$-$\phi$ and $T$-$A$ solutions.}
		\label{Nvariation_field}
	\end{figure}

	In Fig.~\ref{Nvariation_field}, the profile of ${h}_t$ for various $N$ values in the TS model are shown. The solution with $N=1$ gives a linear distribution of ${h}_t$ across the thickness of the tape, which is similar to the profile obtained with the $T$-$A$-formulation. When $N$ increases, the solution of the TS model approaches the \mbox{$H$-$\phi$} solution.  With $N=6$, the TS solution is very close to the reference one. Note that in this case the value of~${h}_t$ in the middle of the tape is comparable to the value of $h_n$ at its extremities (see Fig.~\ref{MagneticField_Width_packed} and ~\ref{Nvariation_field}). The tangential components certainly have an determining impact when modeling multiple tapes in configurations similar to this example.
	
	\begin{figure} [t]
		\centering	
		\includegraphics[]{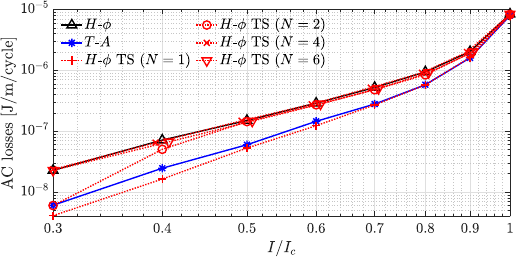}
		\label{Nvariation_losses}
		\caption{ Total AC losses as a function of the normalized transport current for the closely-packed tapes example. The solution of the TS model with different $N$ values is compared with the $H$-$\phi$ and \mbox{$T$-$A$} solutions. Note the convergence of the TS solution to the $H$-$\phi$ at low current transport. The increase in $N$ permits the TS to take top/bottom losses into account.}
		\label{Nvariation}
	\end{figure}
	
	The AC losses as a function of $I/I_c$ obtained with the three models are also presented in~Fig.~\ref{LossesComp}. In the TS model, $N=11$ was considered for the sake of comparison with the reference solution and with the single tape example presented in Section~\ref{Onetape}. The TS solution perfectly agrees with the $H$-$\phi$ solution. The $T$-$A$-formulation underestimates the AC losses at low transport current. Indeed, with a linear variation of the fields inside the tape, so-called top/bottom losses are not taken into account by the $T$-$A$-formulation.
	
	In order to define a minimum number of 1-D elements necessary to obtain an accurate solution in terms of AC loss in the TS model, simulations were performed using different $N$ and $F_c$ values. The AC losses as a function of $F_c=I/I_c$ for $N=1$, 2, 4 and 6 are presented in \mbox{Fig.~\ref{Nvariation}}. With $N=1$, $h_t$ does not fully penetrates the tape, and the TS model also underestimates the AC losses at low transport current. With the increase in~$N$, one observes the convergence of the TS solution towards the $H$-$\phi$ solution. 
	
	The number of DoFs, the CPU time and the total number of time iterations (NoIs) for the two closely-packed tapes example {with $I/I_c=0.9$} are summarized in Table~\ref{TableDoFs}. The proposed TS model with $N=1$ represented a reduction of more than 40\% in the total number of DoFs and was more than seven times faster than the reference $H$-$\phi$ model. The three parameters in Table~\ref{TableDoFs}  increased with $N$. Even with $N=11$, the reduction in the number of DoFs in the TS model is still about 30\%, and simulations are 1.3 faster than the $H$-$\phi$ model with complete discretization of the tape. The application of the TS model with $N=4$ shows a good compromise between solution accuracy and computational cost even for the case of closely packed HTS tapes with anti-parallel transport current. 
	
	\begin{table}[t]
		\caption{ Number of DoFs, CPU time, and total number of iterations (NoIs) in the $H$-$\phi$ and in the TS models for different $N$ values {and $I/I_c=0.9$ for the two closely-packed tapes example}.}
		\begin{scriptsize}
			\begin{tabular}{@{}lcccccc@{}}
				\toprule
				& $H$-$\phi$ & \begin{tabular}[c]{@{}c@{}}TS\\ ($N=1$)\end{tabular} & \begin{tabular}[c]{@{}c@{}}TS\\ ($N=2$)\end{tabular} & \begin{tabular}[c]{@{}c@{}}TS\\ ($N=4$)\end{tabular} & \begin{tabular}[c]{@{}c@{}}TS\\ ($N=6$)\end{tabular} & \begin{tabular}[c]{@{}c@{}}TS\\ ($N=11$)\end{tabular} \\ \midrule
				DoFs & 33188      & 18727                                                & 19127                                                & 19927                                                & 20727                                                & 22727                                                 \\
				Time & 4015.65s   & 553.57s                                              & 563.34s                                              & 806.30s                                              & 1252.21s                                             & 3079.08s                                              \\
				NoIs  & 488        & 114                                                  & 114                                                  & 125                                                  & 188                                                  & 383                                                   \\ \bottomrule
			\end{tabular}
		\end{scriptsize}
		\label{TableDoFs}
	\end{table}
	
	With the $T$-$A$ model, the number of DoFs in this example was 33282 and the solution was obtained after 141\,s and 481 iterations. This number of DoFs is comparable to the \mbox{$H$-$\phi$ model} since second-order finite elements are used for~$A$. Despite this, the CPU time was shorter with the $T$-$A$ than the proposed TS model with $N=1$. This difference may be related to the solvers: the \mbox{$T$-$A$-formulation} is implemented in Comsol, and the two other models are in Gmsh/GetDP. A fair comparison in terms of computational costs would require these models to be implemented in the same software environment. 
	
	\section{Realistic Examples}
	
	\subsection{Racetrack Coil}
	
	To demonstrate the applicability of the proposed model for modeling large-scale HTS devices, we considered a stack of twenty HTS tapes in a \mbox{2-D infinitely} long representation of a racetrack coil. The tapes have the same physical and geometrical characteristics as in the two previous examples and an inter-tape separation of $L=250$\,$\mu$m. A transport current of $0.9I_c$ at a frequency of~$f=50$\,Hz was imposed in each tape in two configurations: (i) all the tapes carrying the same transport current and (ii) each subsequent pair of tapes carrying anti-parallel currents of equal magnitude.
	
	According to the results presented in the last example, \mbox{$N=4$} in the TS model is enough to provide a good estimation of the AC losses in the tapes. Therefore, the results obtained with the TS model with $N=4$ for the racetrack coil problem were compared to the results from the $H$-$\phi$-formulation in terms of local field distribution and AC losses. Each tape was represented as as a reduced-dimension geometry and modeled using the proposed TS model.
	
\begin{figure} [t]
			\centering	
		\begin{subfigure}{.22\textwidth}
			\includegraphics[width=1\textwidth]{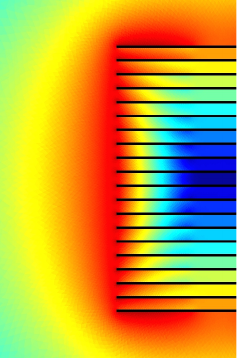}
			\caption{\small $H$-$\phi$ formulation}
		\end{subfigure}
		\begin{subfigure}{.22\textwidth}
			\includegraphics[width=1\textwidth]{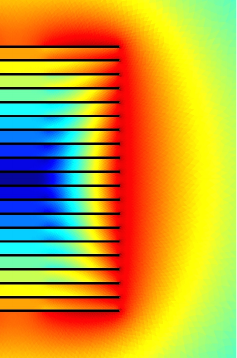}
			\caption{\small $H$-$\phi$ TS model}
		\end{subfigure} \\ 
		\begin{subfigure}{.5\textwidth}
			\vspace{1.5mm}
			\hspace{-2mm}
			\includegraphics[]{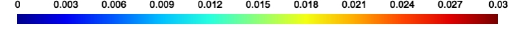}
			\label{Induction_Racetrackcoil_direct_label}
		\end{subfigure}
		\caption{ Norm of magnetic flux density $|\bm{b}|$~[T] for the racetrack coil example with twenty HTS tapes in a infinitely long 2-D representation at $t=T/4$. A transport current of \mbox{$I=0.9I_c$} is imposed in each tape. Half of the domain is presented due to the symmetry. (a) Reference solution from the $H$-$\phi$-formulation, and (b) solution from the TS model with \mbox{$N=4$.}}
		\label{RacetrackCoilSame}
	\end{figure}

\begin{figure} [t]
		\centering	
		\begin{subfigure}{.22\textwidth}
			\includegraphics[width=1\textwidth]{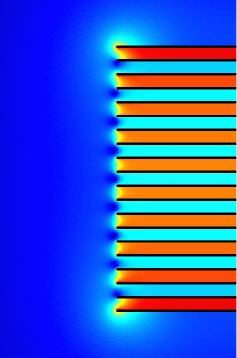}
			\caption{\small $H$-$\phi$ formulation}
		\end{subfigure}
		\begin{subfigure}{.22\textwidth}
			\includegraphics[width=1\textwidth]{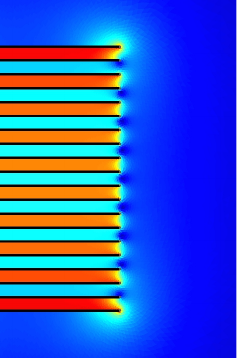}
			\caption{\small $H$-$\phi$ TS model}
		\end{subfigure} \\ 
		\begin{subfigure}{.5\textwidth}
			\vspace{1.5mm}
			\hspace{-2mm}
			\includegraphics[]{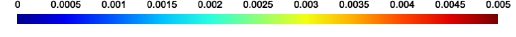}
		\end{subfigure}
		\caption{  Norm of magnetic flux density $|\bm{b}|$~[T] for the racetrack coil example with twenty HTS tapes in a infinitely long 2-D representation at $t=T/4$. An anti-parallel transport current of $I=0.9I_c$ is imposed in each subsequent pair of tapes. Half of the domain is presented due to the symmetry. (a) Reference solution from the $H$-$\phi$-formulation, and (b) solution from the TS model with \mbox{$N=4$.}}
		\label{RacetrackCoilAnti}
	\end{figure}

	Fig.~\ref{RacetrackCoilSame} shows the norm of the flux density at $t=T/4$ in the case where the same transport current is imposed in each tape. Fig.~\ref{RacetrackCoilAnti} shows the equivalent distribution for the case with anti-parallel currents imposed in each pair of tapes. In both figures, the reference solution is presented on the left and the TS model solution is presented on the right. Excellent agreement is observed in all cases.
	
\begin{figure}[t]
		\centering
		\includegraphics[]{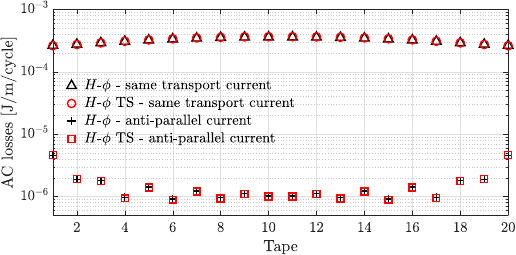}
		\caption{\small Total AC losses in each tape of the racetrack coil. Results are presented for the two studied cases: same transport current in each tape and anti-parallel currents in each pair of tapes.}
		\label{RacetrackCoil_losses}
		\vspace{4mm}
	\end{figure}

	In Fig.~\ref{RacetrackCoil_losses}, we present the AC losses per cycle as a function of the position of the tapes, with 1 corresponding to the bottom tape and 20 to the top tape. Note that the AC losses computed with the TS model fit perfectly the losses obtained with the $H$-$\phi$ reference model. This demonstrates that the TS model with the appropriate number of 1-D elements in the thin film representation of the tape is suitable for modeling HTS systems involving multiple tapes.
	
	\subsection{HTS Tape with ferromagnetic substrate}
\begin{figure}[t]
		\begin{subfigure}{0.24\textwidth}
			\includegraphics[]{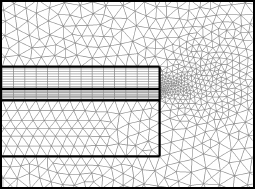}
			\caption{Full mesh}
			\label{HTStapeFull}
		\end{subfigure}\hspace{0.5mm}
		\begin{subfigure}{0.24\textwidth}
			\includegraphics[]{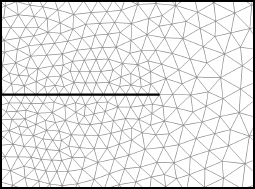}
			\caption{TS model mesh}
			\label{HTStapeTS}
		\end{subfigure} 
		\caption{ Zoom of the meshes near the right edge of the tape: (a) full mesh for standard FE application (reference model), and (b) simplified mesh used with the TS model. Although all layers of the HTS tape are represented as a single line in the TS model, the stacking order is still respected in the virtual discretization.}
		\label{HTStape}
		\vspace{4mm}
	\end{figure}

	\vspace{3mm}
	In our last example, we considered one HTS tape comprising a stack of three thin layers: (i) the bottom layer is a ferromagnetic substrate ($\sigma_{\text{subs}}=10$\,kS/m, $\mu_{r,\text{subs}}=100$, $d_\text{subs}=50$\,$\mu$m), (ii) the middle layer was the HTS ($\mu_{r,\text{HTS}}=1$, $d_\text{HTS}=10$\,$\mu$m), and (iii) the top layer was a silver stabilizer ($\sigma_\text{Silver}=63$\,MS/m, $\mu_{r,\text{Silver}}=1$, $d_\text{Silver}=20$\,$\mu$m). As in the previous examples, the thicknesses of the HTS and silver layers, $d_\text{HTS}$ and $d_\text{Silver}$, respectively, were scaled up to ensure reasonable accuracy of the reference solution. In the TS model, the three layers were collapsed on a single surface. Fig.~\ref{HTStape} shows the geometry and mesh simplifications in the TS model (Fig.~\ref{HTStapeTS}) compared with the reference geometry (Fig.~\ref{HTStapeFull}).
	
\begin{figure}[t]
			\begin{subfigure} {0.24\textwidth}
			\centering
			\includegraphics[]{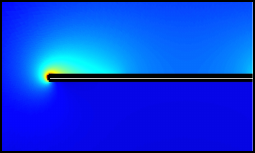}
			\caption{ $H$-$\phi$ formulation ($t= T/8$)}
			\vspace{0mm}
		\end{subfigure}
		\begin{subfigure} {0.24\textwidth}
			\centering
			\includegraphics[]{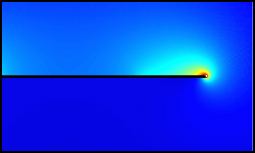}
			\caption{ $H$-$\phi$ TS model ($t=T/8$)}
			\vspace{0mm}
		\end{subfigure}
		\begin{subfigure}{0.49\textwidth}
			\vspace{0.1cm}
			\includegraphics[]{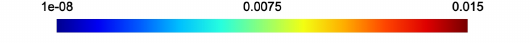}
		\end{subfigure}
		\vspace{-3mm}
		\caption{  Norm of magnetic flux density $|\bm{b}|$~[T] for a single HTS tape comprising three thin layers with $I/I_c=0.9$ at $T/2$. (a) Fully discretized \mbox{$H$-$\phi$} reference model, and (b) proposed TS approach with $N=6$.}
		\label{MagFluxTape}
	\end{figure}

	We initially considered the TS model with one virtual element in the substrate and the silver layers, and four virtual elements in the HTS layer (i.e., $N=6$ in total). A high dense mesh inside the tape was considered in the reference model (Fig.~\ref{HTStapeFull}). Fig.~\ref{MagFluxTape} shows the magnetic flux density magnitude in the surroundings of the tape with $I/I_c=0.9$ and $f=50$\,Hz at $t=T/2$. Due to the high permeability of the substrate, the magnetic flux density is absorbed by this layer and magnified near the top face of the tape. The TS model well represents this behavior, and the solution visually agrees with the reference one.
	
\begin{figure}[t]
			\begin{subfigure}{0.24\textwidth}
			\includegraphics[]{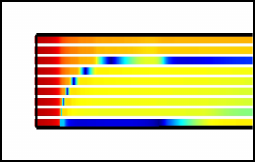}
			\caption{$H$-$\phi$ formulation ($t=T/2$)}
			\label{currentFullT}
		\end{subfigure}\hspace{0.5mm}
		\begin{subfigure}{0.24\textwidth}
			\includegraphics[]{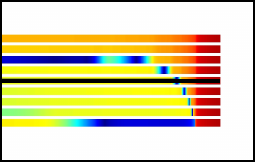}
			\caption{$H$-$\phi$ TS model ($t=T/2$)}
			\label{currentThinT}
		\end{subfigure}
		\begin{subfigure}{0.49\textwidth}
			\vspace{0.1cm}
			\includegraphics[]{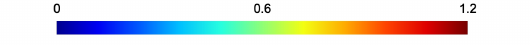}
		\end{subfigure}
		\vspace{-2mm}
		\caption{ Normalized relative current density ($|\bm{j}_z|/j_c$) near the extremity of the HTS tape with three layer at $t=T/2$: (a) $H$-$\phi$ reference solution, and (b) $H$-$\phi$ TS solution with $N=11$. The TS solution was obtained by a projection of the $|\bm{j}_z|/j_c$ on the virtual position of the 1-D elements. The black line in the middle of (b) represents the tape in the TS model (a line in 2-D).}
		\label{CurrentDensityTape}
	\end{figure}

	Next, we increased the number of virtual elements used to represent the HTS layer in order obverse the uneven distribution of the current density in the HTS layer. We considered nine elements across the thickness of the HTS layer, and one element in the substrate and the silver layers ($N=11$). Fig.~\ref{CurrentDensityTape} shows the normalized current density $|\bm{j}_z|/j_c$. Note the differences between this distribution and the one when only the HTS layer was modeled (Fig.~\ref{Current}). Under this configuration, the current is concentrated near the silver layer, i.e. far from the substrate.
	
	Simulations with the TS model were twice as fast as the simulations using the reference model with same size mesh elements on the surfaces of the tapes. The number of DoFs was reduced by around 30\% in this case. However, as discussed in this paper,  a coarser mesh could have been used in the TS model. The considered high-density mesh served to validate the TS model in terms of accuracy.
	
\begin{figure}[t]
		\centering
		\includegraphics[]{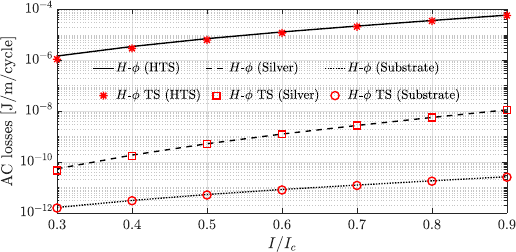}
		\caption{ Total AC losses in each layer of an HTS tape as a function of the relative transport current $I/I_c$. Note the good agreement between the solutions in each layer of the HTS tape.}
		\label{Tape_losses}
	\end{figure}

	The AC losses per cycle of time as a function of the transport current in each layer of the tape are presented in Fig.~\ref{Tape_losses}. Given the low resistivity value of HTSs, the current flows mainly in this layer. Therefore, AC losses in the HTS layer are more important than in the other two layers. This example demonstrates that the TS approach can model various HTS devices architectures including HTS tapes comprising multiple layers made of different materials.
	
	\section{Conclusion}
	
	In this paper, a new thin shell (TS) model for modeling thin HTS tapes in 2-D was developed and validated through examples. The model is based on the $H$-$\phi$-formulation and the addition of auxiliary 1-D FE equations for the variation of the tangential components of the magnetic field inside the tapes. This procedure allows representing the nonlinearities associated with the HTS material in time-transient FE simulations. Compared to a reference solution based on standard FE with the \mbox{$H$-$\phi$-formulation,} the TS model provided good results for all the studied cases while also reducing the number of DoFs by more than $40\%$ and speeding up simulations by seven times in some cases.

	The TS model was also compared with the widely used \mbox{$T$-$A$-formulation} in terms of local accuracy and AC losses. For problems where the magnetic field normal components are more intense than the tangential components, the TS model with $N=1$ is comparable to the \mbox{$T$-$A$-formulation} since both models consider a linear field distribution across the thickness of the tape. However, when the tapes are closely-packed together, the \mbox{$T$-$A$-formulation} does not provide good solutions. The TS model application with $N>1$ has shown to give a correct solution in these cases. We demonstrated that the virtual discretization can be especially interesting to tackle problems involving multiple thin layers, such as those in real HTS tapes. 
		
	{However, we must emphasize that the TS model is very general and its use goes much beyond modeling HTS systems.} Indeed, the presented model represents a generalization of the classical TS model with interface conditions (ICs) defined as 1-D auxiliary problems across the thickness of the thin region. This allows tackling problems involving multilayers of different materials and the use of special basis functions in the 1-D problem. In addition, the presented methodology can easily be extended to a pure $H$ or \mbox{$A$-formulation} with the ICs properly defined. Its application in a $A$-formulation would be relevant for modeling nonlinear ferromagnetic thin regions in time-transient analysis, such as nonlinear shielding problems.
	
	In future work, the presented TS model will be extended for 3-D simulation of HTS devices with transport current and external field excitations.
		
	\section*{Acknowledgments}
	\par The authors would like to acknowledge Fr\'ed\'eric Trillaud for sharing his codes on this matter, Ruth V. Sabariego and Christophe Geuzaine for their insights concerning this work and the help with the FE implementation using Gmsh and GetDP, and Alexandre Arsenault for all the discussions concerning the $H$-$\phi$-formulation.
	
	\par This work has been supported by the Coordination for the Improvement of Higher Education Personnel (CAPES) - Brazil - Finance Code 001, and the Fonds de Recherche du Québec - Nature et Technologies (FRQNT).

	\ifCLASSOPTIONcaptionsoff
	\newpage
	\fi
	
	\bibliographystyle{IEEEtran}
	\bibliography{IEEEabrv,Bibliography}

\end{document}